\documentclass[prb,aps,twocolumn,draft,tightenlines,%
showpacs,floatfix,superscriptaddress]{revtex4}

\usepackage{graphicx}
\usepackage{dcolumn}
\usepackage{bm}
\usepackage{psfig}
\begin{document}

\title{Spin-dependent quantum transport in  periodic magnetic 
modulations: Aharonov-Bohm ring structure as a spin filter}
\author{M. W. Wu}
\thanks{Author to whom correspondence should be addressed}%
\email{mwwu@ustc.edu.cn}%
\affiliation{Hefei National Laboratory for Physical Sciences at
  Microscale, University of Science and Technology of China, 
Hefei, Anhui, 230026, China}
\affiliation{Department of Physics, University of Science and
Technology of China, Hefei, Anhui, 230026, China}
\altaffiliation{Mailing Address.}
\author{J. Zhou}
\affiliation{Department of Physics, University of Science and
Technology of China, Hefei, Anhui, 230026, China}
\author{Q. W. Shi}
\affiliation{Hefei National Laboratory for Physical Sciences at
  Microscale, University of Science and
Technology of China,  Hefei, Anhui, 230026, China}

\date{\today}

\begin{abstract}
Quantum interference in Aharonov-Bohm (AB) ring structure
provides additional control of spin at mesoscopic scale. 
We propose a scheme for spin filter by studying the 
coherent transport through the AB  structure
with lateral magnetic modulation on both arms of the ring structure.
Large spin polarized current can be obtained with many energy channels. 
\end{abstract}
\pacs{85.75.-d, 73.23.Ad, 72.25.-b}
\maketitle

Spin-based devices hold promises for future applications in
conventional as well as in quantum computer 
hardware.\cite{prinz,loss,spin}
Spin injection across interfaces is one of the crucial ingredients for
such applications. However, an efficiency of spin injection 
through ideal ferromagnet/non-magnetic semiconductor interfaces
is disappointingly small due to the large conductivity mismatch\cite{schmidt}
between the magnetic ferromagnet and the semiconductor. The
use of spin filters\cite{filter}
is therefore an alternative approach which can significantly 
enhance spin injection efficiencies. 
In most of these works, spin-selective  barriers or stubs\cite{stub} 
are essential to realize the spin polarization (SP).
Recently  we propose a  scheme for spin filters
that the SP is generated during the transport through quantum wires
with periodic lateral magnetic modulation which is much {\em weaker}
than the Fermi energy $E_f$ of the leads (Electrons therefore 
transport without tunneling
through any barrier or being mode-selected by any stub).
100\ \% SP  through the filter is predicted which origins
from the mismatch of the effective spin-band-gaps induced by the
Bragg-diffraction from the periodic modulation.\cite{wu}
The spin current density is up to 5.45\ nA.
However there is only one fixed energy 
interval for such SP.

In this letter we propose an improved design of the
spin filter in the Aharonov-Bohm (AB) ring structure,\cite{ab}
an AB frame, coupled symmetrically to two leads
to further increase the SP (by one order of magnitude) and the energy 
intervals (channels) for the SP by the additional
degree of freedom, {\em ie.}  the AB flux.
A schematic of the spin filter  is shown in Fig.\ 1. 
A periodic spin dependent modulation with Zeeman-like form
 $V_\sigma(x,y)=\sigma V_0g(x,y)$ is applied symmetrically along both arms 
of the microstructure.
Here $g(x,y)=1$ if $(x,y)$ locates at the gray areas (A layers), and 
0 otherwise.  $\sigma$ is $\pm1$ for spin-up and -down electrons
respectively. $V_0$ denotes a spin-independent parameter for 
the strength of the potential. This modulation can be 
realized by sticking magnetic strips on top of the sample or
using magnetic semiconductor as the A layer.
For $E_f\gg V_0$, spin-up and -down electrons experience different 
potentials: the spin-up electrons coherently transport under the modulation 
of the ``transparent'' barriers on the 
arms while the spin-down ones do under the modulation of wells.  
The AB flux through the AB frame is introduced by a 
homogeneous magnetic field $B$. For the sake of simplicity
for the experimental realization, we assume this magnetic field is applied not
only inside the frame, but also on the arms.

\begin{figure}[htb]
\vskip-0.1cm
\centerline{\psfig{figure=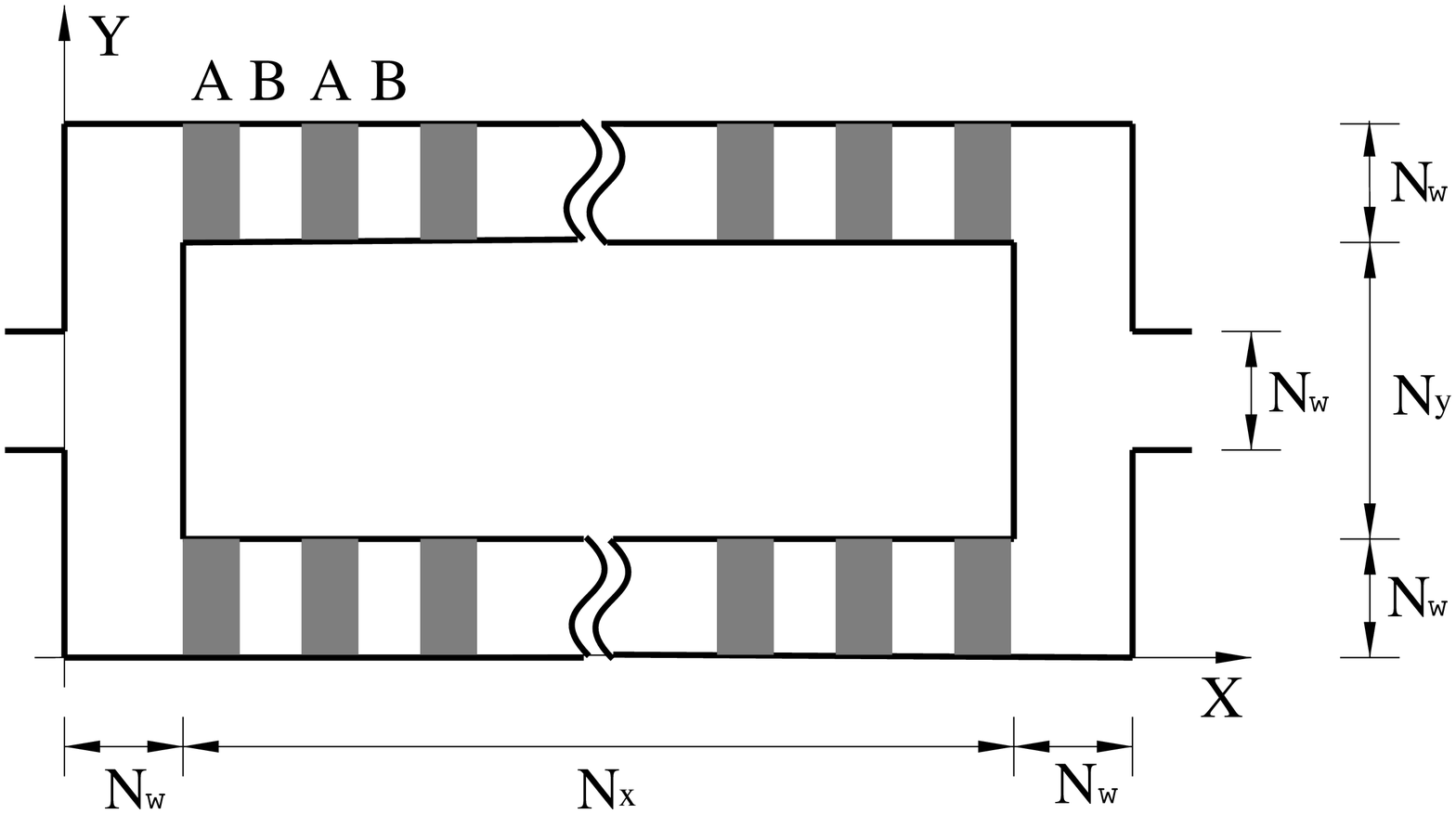,width=7.5cm,height=3cm,angle=0}}
\vskip-0.3cm
  \caption{Schematic of the spin filter in AB ring structure.}
\end{figure}

We describe the AB frame by a tight-binding Hamiltonian with the nearest-neighbour
approximation:
\begin{eqnarray}
H&=&\sum_{l,m,\sigma}( \frac{\epsilon_{l,m,\sigma}}{2} c^\dagger_{l,m,\sigma}
c_{l,m,\sigma} + 
t_{l,m+1;l,m}c^\dagger_{l,m+1,\sigma}c_{l,m,\sigma} \nonumber \\
&&\mbox{}+ t_{l+1,m;l,m}c^\dagger_{l+1,m,\sigma}c_{l,m,\sigma} +\mbox{H.C.})\ ,
\end{eqnarray}
in which $l$ and $m$ denote the  coordinates along the  $x$- and $y$-axis
respectively.  
$\epsilon_{l,m,\sigma}=\epsilon_0+\sigma V_0$ ($=\epsilon_0$)
when $(l,m)$ locates at the gray (blank) areas of the frame, denotes
the on-site energy with $\epsilon_0=-4t_{0}$.
 $t_{0}=-\hbar^2/(2m^\ast a^2)$ is the hopping energy with $m^\ast$ 
and $a$ standing for the effective mass and  the ``lattice'' constant
respectively. With the vector potential ${\bf A}$ in the Landau gauge,
{\em ie.}, ${\bf A} =(-\frac{1}{2}By,\frac{1}{2}Bx,0)$, the hopping energy between
${\bf r}_i$ [$=(l_i,m_i,0)]$ and ${\bf r}_j$ is given by
$t_{{\bf r}_{i},{\bf r}_{j}}=t_{0}e^{[ie{\bf A}\cdot({\bf r}_{i}-
{\bf r}_{j})/\hbar]}$.

The spin dependent conductance is calculated using  the
Landauer-B\"{u}ttiker\cite{Bu} formula with the help of 
the Green function method.\cite{Da} The two-terminal spin-resolved 
conductance is given by
$G^{\sigma \sigma^\prime}=(e^2/h)\mbox{Tr}[\Gamma^{\sigma}_{1}
G^{\sigma\sigma^\prime+}_{1(2N_w+N_x)}
\Gamma^{\sigma^\prime}_{2N_w+N_x}G^{\sigma
^\prime\sigma -}_{(2N_w+N_x)1}]$ with  $\Gamma_1$ 
($\Gamma_{2N_w+N_x}$) representing the
self-energy  function for the isolated ideal leads.\cite{Da} 
We choose the  perfect ideal ohmic contact between the
leads and the semiconductor. $G^{\sigma\sigma^\prime+}_{1(2N_w+N_x)}$ and 
$G^{\sigma\sigma^\prime-}_{(2N_w+N_x)1}$ are the 
retarded and advanced Green functions
for the conductor, but with the effect of the leads included. 
The trace is performed over the spatial degrees of freedom along the 
$y$-axis. The spin dependent current within an energy window
$[E,E+\Delta E]$ is given by $I_\sigma=\int_{E}^{E+\Delta
  E}G^{\sigma\sigma}(E)dE$.

We perform a numerical calculation for an AB frame with 
fixed arm width $N_w a=20a$.  A hard wall potential is applied in this
transverse direction which makes the lowest energy of the
$n$th subband (mode) be $\epsilon_n=2|t_0|\{1-\cos[n\pi/(N_w+1)]\}$.
The total length and width of frame are
$500a$ and $90a$ 
respectively with $a=9.53$\ {\AA} throughout the computation.
The length of A layer $L_{A}$ is taken to be the same as that of
B layer with $L_A=L_B=8a$.  We take the Zeeman splitting energy
$V_0=0.001|t_0|$. Unless otherwise specified, $B=0.45\phi_{0}/S$ with 
$S=(2N_w+N_x)(2N_w+N_y)a^2$ 
denoting the area of the AB frame and
$\phi_{0}=h/e$ standing for the quantum unit of flux. It is noted that
$g\mu_BB\ll V_0$. 

\begin{figure}[htb]
\vskip-0.3cm
  \psfig{figure=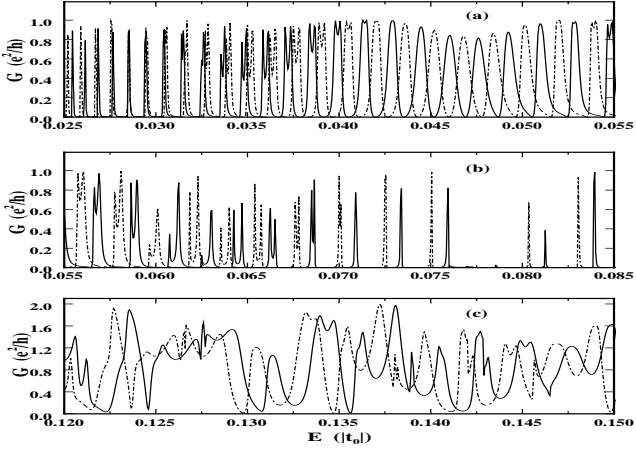,width=9cm,height=6.5cm,angle=0} 
\vskip-0.3cm
  \caption{Conductance versus the energy of the electron with (a) and (b) compose the first
  subband and (c) gives part of the second subband. Solid curve: $G^{\uparrow\uparrow}$;
Chain curve:  $G^{\downarrow\downarrow}$.}
\end{figure}

In Fig.\ 2 (a) and (b), the conductance is plotted as a function of the Fermi energy $E$ 
of the leads, with $E$ limited to the first subband.
It is seen from the figure that for the AB ring structure with periodic
magnetic modulations, due to the
interference  many energy windows are opened which give large SP,
in contrast to the quantum wire where only {\em one} energy window is 
opened for large SP.\cite{wu} The largest spin current density can be
obtained from the energy window $[0.0566|t_0|,0.0576|t_0|]$ with 
$I^{SP}_{\uparrow}=I_{\uparrow}-I_{\downarrow}\approx 16.8$\ nA  for 
spin-up current and from the window $[0.0556|t_0|,0.0565|t_0|]$ with
$I^{SP}_{\downarrow}=I_{\downarrow}-I_{\uparrow}\approx 13.4$\ nA 
for spin-down current. A zero conductance of spin-up electron 
near $E=0.06|t_0|$ 
corresponds to the energy gap due to the modulations predicted in 1D case, 
with the wave length of $\lambda\sim 2(L_{A}+L_{B})$.\cite{wu}
A spin-independent gap between $0.075|t_0|$ to $0.08|t_0|$ 
origins from the interference effect of the four rectangular 
bends of the AB frame.
By reshaping these four bends, we obtain the shift of this gap. 
Larger spin-current density can be obtained with multi-mode
transport. In Fig.\ 2(c)
we plot the conductance versus the energy in the regime of
 the second subband and
therefore two modes participate the carrier transport. 
By choosing the energy window $[0.1217|t_0|,0.1234|t_0|]$, 
one gets the spin current density
$I^{SP}_{\downarrow}=43.6$\ nA.

\begin{figure}[htb]
\vskip-0.3cm
\psfig{figure=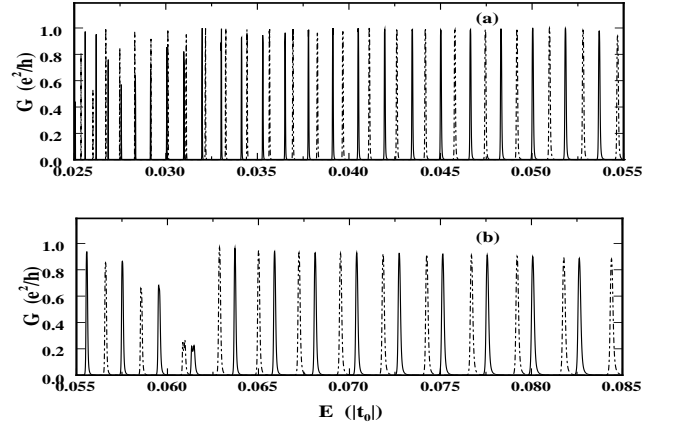,width=9cm,height=6.5cm,angle=0}
\vskip-0.3cm
\caption{Conductance versus the energy of the electron from the 1D model. 
 Solid curve: $G^{\uparrow\uparrow}$;
Chain curve:  $G^{\downarrow\downarrow}$.}
\end{figure}

In order to further elucidate above features of the conductance,
we simplify the AB frame by taking $N_w=1$. Therefore
one can solve the problem analytically with the  approach developed by 
Xiong.\cite{xiong} The Hamiltonian, including the leads can now be 
simplified as 
\begin{eqnarray}
H&=&\sum_{N\ge i\ge 1\sigma}
\epsilon_{\mu ,i,\sigma}/2 c_{\mu ,i,\sigma}^{\dagger}c_{\mu ,i,\sigma}
+\sum_{\mu\sigma} \Big[ t_{0}(c_{\mu,1,\sigma}^{\dagger}c_{L_{0},\sigma}\nonumber\\
&&\mbox{}\hspace{-0.6cm}+c_{R_{0},\sigma}^{\dagger}c_{\mu,N,\sigma})
+\sum_{N>i\ge
  1\sigma}t_{\mu,i+1;\mu,i}c_{\mu,i+1,\sigma}^{\dagger}c_{\mu,i,\sigma}\Big] 
\nonumber\\
&&\mbox{}\hspace{-0.6cm}+ t_{0}\sum_{j>0\sigma}(c_{L_{-j+1},\sigma}^{\dagger}c_{L_{-j},\sigma}+
c_{R_{j},\sigma}^{\dagger}c_{R_{j-1},\sigma})+\mbox{H.C.}\ ,
\end{eqnarray}
where $\mu=1$, 2 denotes two arms and $i=1\cdots N$ stands for the
index of sites  on each arm.
$L_{j}$ ($R_{j}$) labels the site of left (right) lead.  
Site index increases from the left to the right and 
two junctions between the leads and the arms are defined as the
end of the leads denoted by $L_{0}$ and $R_{0}$. 
We take the one-dimensional (1D) 
AB frame as the outer edge of the 2D structure in Fig.\ 1.
Similar to the 2D problem, $\epsilon_{\mu,i,\sigma}=\epsilon^{\prime}_{0}
+\sigma V_{0}$ ($\epsilon^{\prime}_{0}$) 
when $i$ locates  at A layer (otherwise). We take
$\epsilon^{\prime}_{0}=\epsilon_{1}+2|t_{0}|$ where $\epsilon_{1}$ 
is the lowest energy of the first subband  
and denote $E_{\mu,i,\sigma}=E-\epsilon_{\mu,i,\sigma}$ with $E$ 
being the energy of the  incident electron.
The wave-function  can be expressed as\cite{xiong}
$|\psi\rangle=[\sum_{j\ge 0,\sigma}(A_{L_{-j},\sigma}c_{L_{-j},\sigma}^\dagger
+A_{R_{j},\sigma}c_{R_{j},\sigma}^\dagger)
+\sum_{\mu,N\ge i\ge 1,\sigma}B_{\mu,i,\sigma}c_{\mu,i,\sigma}^\dagger]
|0\rangle$, with  the coefficients
$A_{L_{j},\sigma}$, $A_{R_{j},\sigma}$ and $B_{\mu,i,\sigma}$ 
being determined from the Schr\"{o}dinger equation 
$H|\psi\rangle=E|\psi\rangle$. By assuming 
a plane wave with unity amplitude and 
spin $\sigma$ injected from the left lead, one has 
$A_{L_{j},\sigma}=
e^{ikj}+r_{\sigma}e^{-ikj}$ and $A_{R_{j},\sigma}=t_{\sigma}e^{ikj}$
with $r_{\sigma}$ and $t_{\sigma}$ 
standing for the reflection and transmission amplitudes 
 and $k$ being the wave-vector
satisfying $2t_{0}\cos{k}=E-\epsilon^{\prime}_{0}$.
The transmission coefficient $T_{\sigma}=|t_{\sigma}|^{2}$
is therefore given as 
\begin{equation}
\label{t}
t_{\sigma}=\frac{-4i\sin k\cos(\frac{\phi}{2})
  C_{\sigma}}{(2C_{N,\sigma}+e^{-ik})(2D_{N,\sigma}+e^{-ik})
-4\cos^{2}(\frac{\phi}{2})C_{\sigma}D_{\sigma}}\ ,
\end{equation}
in which  $C_{i,\sigma}=(E_{i,\sigma}-C_{i-1,\sigma})^{-1}$ with
$C_{1,\sigma}=E_{1,\sigma}^{-1}$ and $D_{i,\sigma}=
(E_{N-i+1,\sigma}-D_{i-1,\sigma})^{-1}$ with
$D_{1,\sigma}=E_{N,\sigma}^{-1}$. 
$C_{\sigma}=\prod_{i=1}^{N}C_{i,\sigma}$ and 
$D_{\sigma}=\prod_{i=1}^{N}D_{i,\sigma}$.
\begin{figure}[htb]
\vskip-0.3cm
\centerline{\psfig{figure=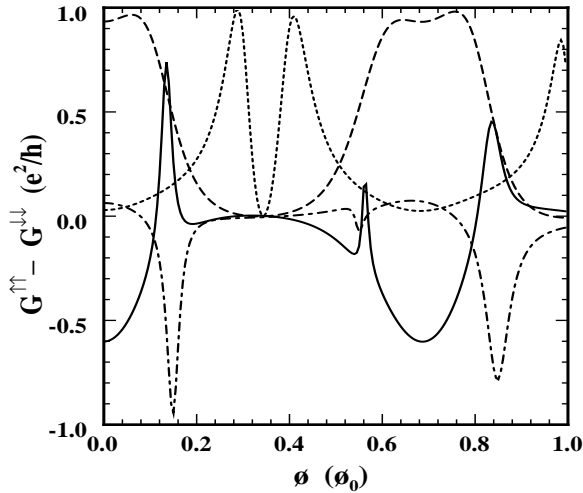,width=8cm,height=7cm,angle=0}}
\vskip-0.1cm
\caption{$G^{\uparrow\uparrow}-G^{\downarrow\downarrow}$ versus the AB flux $\phi$
for different energies.
Solid curve: $E=0.0585|t_{0}|$; Dotted curve: $0.0589|t_{0}|$; Dashed curve:
  $0.0592|t_{0}|$; Chain curve: $E=0.0595|t_{0}|$.}
\end{figure} 

We apply the same magnetic field (and therefore the AB flux) and
use the same parameters as in the 2D case.
The spin-dependent conductances $G^{\sigma\sigma}=(e^2/h)T_\sigma$
calculated from Eq.\ (\ref{t}) are plotted in Fig.\ 3 and exhibit similar
behavior as the 2D case in Fig.\ 2, including the band gap
due to the magnetic modulation near  $E=0.06|t_0|$. 
However, the peaks of 1D structure is much narrower and 
the gap caused by the four rectangular bends in Fig.\ 2
disappears as the bends do not exist in the 1D problem.
  
Finally we consider the effect of AB flux. In Fig.\ 4, 
the difference of the conductances of spin-up and -down electrons 
of the 2D AB frame 
 $G^{\uparrow\uparrow}-G^{\downarrow\downarrow}$
is plotted as a function of the flux $\phi$ for four typical enegeies: 
$E=0.0585|t_{0}|$ (solid curve), $0.0589|t_{0}|$ 
(dotted curve), $0.0592|t_{0}|$ (dashed curve)
and $0.0595|t_{0}|$ (chain curve). One finds efficient modulation due to the 
AB flux. Moreover, as the magnetic field is applied also on the 
arms of the AB  frame, it is seen from the figure that the 
period is now roughly about $0.7\phi_{0}$, in contrast to $\phi_0$ for the case
when the flux is confined inside the frame.

In summary, we have proposed a scheme for spin filter 
by studying the coherent transport through an AB-ring structure with
lateral magnetic modulation. Large SP is predicted and is shown 
to be accessible with many energy intervals.
The magnetic modulation can be realized by sticking the magnetic 
strips on top of the sample or using magnetic semiconductor as A layers.

One of the authors (MWW) was supported by the  ``100 Person Project'' of Chinese Academy of
Sciences and Natural Science Foundation of China under Grant Nos. 9030312 and 10247002.
He would like to thank S. J. Xiong for valuable discussions.

\end{document}